*Research Article*

# PV-Powered CoMP-Based Green Cellular Networks with a Standby Grid Supply

**Abu Jahid,[1] Abdullah Bin Shams,[2] and Md. Farhad Hossain[3]**

[1]*Department of Electrical, Electronic and Communication Engineering, Military Institute of Science and Technology, Dhaka 1216, Bangladesh*
[2]*Department of Electrical and Electronic Engineering, Islamic University of Technology, Gazipur 1704, Bangladesh*
[3]*Department of Electrical and Electronic Engineering, Bangladesh University of Engineering and Technology, Dhaka 1000, Bangladesh*

Correspondence should be addressed to Md. Farhad Hossain; mfarhadhossain@eee.buet.ac.bd

This paper proposes a novel framework for PV-powered cellular networks with a standby grid supply and an essential energy management technique for achieving envisaged green networks. The proposal considers an emerging cellular network architecture employing two types of coordinated multipoint (CoMP) transmission techniques for serving the subscribers. Under the proposed framework, each base station (BS) is powered by an individual PV solar energy module having an independent storage device. BSs are also connected to the conventional grid supply for meeting additional energy demand. We also propose a dynamic inter-BS solar energy sharing policy through a transmission line for further greening the proposed network by minimizing the consumption from the grid supply. An extensive simulation-based study in the downlink of a Long-Term Evolution (LTE) cellular system is carried out for evaluating the energy efficiency performance of the proposed framework. System performance is also investigated for identifying the impact of various system parameters including storage factor, storage capacity, solar generation capacity, transmission line loss, and different CoMP techniques.

## 1. Introduction

Due to the recent unprecedented growth in the number of subscribers and diverse data applications, mobile cellular network operators are deploying a higher number of BS in their infrastructure resulting in an exponential increase in energy consumption. Such growth of energy demand in the cellular network industry is exerting enormous detrimental effect on both the economical and the environmental aspects [1–3]. Recent studies suggest that around 50% of the operating expenditure (OPEX) of a cellular system attributes to the energy cost for running the network. On the other hand, it has been reported that the information and communication technology (ICT) industry contributes about 2–2.5% of total carbon emissions and this is expected to increase every year with the exponential growth of mobile traffic [4, 5]. Moreover, this increased energy consumption in cellular networks places heavy burden on the electric grid. Therefore, with the escalating awareness of global warming and energy costs for operating cellular networks, green communications have received considerable attention among the telecommunication researchers and have led to an emerging trend to improve the energy efficiency (EE) of the overall system [4, 6]. Powering the cellular networks as much as possible by renewable energy sources is potentially the best alternative for reducing and even completely phasing out the consumption from the conventional grid supply leading to improved EE and decreased carbon footprint. The most popular among the renewable energy sources are solar, wind, and hydropower. A step towards green communication requires the renewable energy source to be easily integratable with the existing cellular networks. It must also be economical, widely available, and modular and should occupy a smaller area so that it can easily be installed at the vicinity of the BSs. From this perspective, the most feasible and lucrative renewable energy source is photovoltaic (PV) cells.

*1.1. Photovoltaic Power Plants.* As an efficient way to utilize solar energy, PV power plant has received increased attention all over the world due to the fossil fuel crisis and its associated environmental pollution. The tremendous growth of energy consumption around the world has led to an increase in operating cost and global warming. Solar energy offers attractive solutions to reduce carbon footprints and mitigate the global climate change. Burning of nonrenewable energy sources like fossil fuel produces greenhouse gases, whereas PV-based power plants and industries have no such detrimental effect on the environment. Since renewable solar energy is derived from resources which are regenerative, it does not emit carbon. In accordance with the growing trend of PV power plant, solar energy can also be used for a variety of purposes, such as solar-powered BSs for green communications, solar irrigation, solar cold storage, and solar boat systems [7]. Many emerging economies [7–9] have an excellent solar resource and have adopted policies to encourage the development of the solar industry to realize the benefits of PV technology. This can generate positive impact on their economies, as well as on the local and global environment, and improve energy security. Chandel et al. [8] analyzed the potential and cost-effectiveness of solar PV power plant of 2.5 MW for meeting the energy demand of the garment zone in India. A viability analysis [9] of 1 MW PV power plants was conducted in Serbia by taking different types of solar modules to find out the best possibilities of generating high electricity.

*1.2. PV Solar Energy for Green Communications.* BSs in the radio access network (RAN) of cellular mobile networks are the most energy hungry equipment amounting around 60%–80% of the total consumption [10–12], whereas the accumulated energy requirement for user equipment (UE) is around 1% [13]. On the other hand, cellular network data traffic is expected to increase approximately by a factor of ten every five years resulting in a tremendous pressure on energy demand [14]. Thus, this unprecedented growth in energy consumption exerts a detrimental impact on the environment in terms of carbon footprints [4, 5, 15]. Therefore, energy-efficient resource management system in RANs has become the center of focus of the researchers from both academia and industry. This trend has motivated the interest of researchers in an innovative research area called "green communications" concentrating on the environmental effects of cellular networks.

Being inspired for curving down the energy costs, telecom operators have started the deployment of renewable energy sources, such as solar panels, for improving EE of RAN infrastructure. The objective of green cellular communications is to reduce the overall nonrenewable energy consumption leading to improved EE and higher economic benefits. Conventional design approaches focus on optimizing the quality of service (QoS) parameters such as cell coverage, capacity, and throughput with no consideration on the EE aspect of cellular networks. However, with the introduction of green communication paradigm, designed networks must also maintain the same level of QoS while improving EE [11, 16]. On the other hand, the intermittency and spatial randomness of renewable energy generation can severely degrade the system performance of large-scale cellular networks, and hence, it is a fundamental design issue to utilize the harvested energy to sustain traffic demands of users in the network [17]. As a consequence, though renewable energy sources are being deployed in BSs, a provision of conventional grid energy is still required to mitigate for the variability of the renewable energy generation.

Considering the aforementioned concerns, envisioning BSs to be powered by hybrid supplies combining solar energy sources with on-grid sources has become a promising alternative stimulating the proposed work in this paper. In such cellular networks, the primary energy source for BSs is the solar energy. If enough green energy is not available, BSs draw energy from the grid supply for serving its associated UEs. The focus of such green networking is to maximize the usage of solar energy while minimizing the conventional grid energy utilization. The optimal use of solar energy over a period of time depends on the proper energy management techniques integrated into the network operation.

*1.3. Coordinated Multipoint (CoMP) Transmission Technique.* Spectral efficiency (SE) and EE are considered as the prime performance metrics for planning and operation of next-generation cellular networks. SE is a key performance parameter defined as the overall throughput per unit bandwidth. On the other hand, coordinated multipoint (CoMP) transmission has been widely discussed as a promising candidate for future LTE-Advanced (LTE-A) cellular systems [18, 19]. In a cellular network with CoMP, multiple BSs coordinate among themselves for serving a UE in the best possible way. Thus, CoMP has the potential to improve the network performance in terms of interference management, cell-edge throughput, and overall SE as well as EE [20–22]. The downlink CoMP can be categorized into three types based on data availability at multipoint: joint transmission (JT), dynamic point selection (DPS), and coordinated scheduling/coordinated beamforming (CS/CB), which are outlined by 3GPP [20, 23]. In the DPS technique, BS offering the highest SINR is dynamically selected for serving a UE. In contrast, under the JT technique, multiple coordinating BSs transmit data simultaneously to a UE. On the other hand, in the CS/CB technique, signal is transmitted from only one BS by employing beamforming, which is achieved through proper scheduling among the coordinated BSs for avoiding intercell interference.

*1.4. Contributions.* This paper proposes and explores the potential of different approaches for improving the EE

of CoMP transmission-based future cellular networks. The main contributions of this paper can be summarized as follows:

(i) This paper proposes a novel framework for improving the EE of the CoMP-based next-generation cellular networks by employing a hybrid power supply for BSs. Under the proposed framework, PV solar modules work as the primary energy source, while conventional grid power is proposed as the standby source for running BSs in case of insufficient solar energy for serving the UEs with no interruption. The proposed hybrid energy usage scheme is then investigated for both DPS and JT CoMP transmission technique-based cellular networks, which has not been reported yet in literature.

(ii) Then, a technique for maximizing the green energy utilization (i.e., minimizing the consumption from a conventional grid supply) is developed, while BSs are still powered by hybrid sources with the proposed energy usage scheme as outlined above. Therefore, a heuristic policy for sharing green energy (i.e., solar) among the BSs is proposed. For enabling the inter-BS energy transfer, neighboring BSs are proposed to be connected through a resistive lossy transmission line. The proposed energy sharing scheme is also integrated in and investigated for both the DPS and the JT CoMP-based cellular networks. To the best of our knowledge, we are the first to propose a green energy sharing technique for CoMP-based cellular systems.

(iii) Tempo-spatial cellular traffic diversity as well as solar energy generation variability plays a significant role in developing effective green networking techniques. On the other hand, intercell interference, wireless channel propagation model including shadow fading, and BS power consumption model are the other major factors that can affect any system performance. All these factors are taken into consideration in this proposed research and thus makes the network scenario near realistic.

(iv) Extensive simulations are carried out for investigating the energy usage analysis of the proposed framework in terms of various performance metrics such as EE, energy consumption indicator (ECI), and on-grid energy savings. Simulations are carried out considering both temporal traffic diversity over 24 hours and spatial traffic diversity over the entire network area.

(v) The impact of various system parameters including solar storage capacity, storage factor, transmission line loss, solar generation capacity, and CoMP techniques on the performance metrics is demonstrated and critically analyzed. Furthermore, system performance is also compared with that of the existing hybrid non-CoMP-based cellular system having no energy sharing.

The rest of the paper is organized as follows. Section 2 presents a thorough study on the related works. A detailed discussion of the system model along with the network layout and green energy model is outlined in Section 3. In addition, the energy consumption model for macrocell BS and the formulation of performance metrics are also presented in the same section. Section 4 presents the user association policy and the proposed algorithms. Section 5 shows the simulation results with insightful analysis, and finally, Section 6 concludes the paper summarizing the key findings.

## 2. Related Works

Over the last decades, the ever-increasing energy consumption in cellular networks has received intensive attention from regulatory bodies and mobile operators. With the growing awareness of global warming and financial consequences, both researchers and industries have initiated projects to reduce the increasing trend of energy consumption [2, 24, 25]. Switching off some BSs during the low traffic period is the most popular technique for minimizing energy consumption in RAN infrastructure [26–28]. Moreover, the concept of dynamic sectorization of BSs [29] and the traffic-aware intelligent cooperation among BSs [30] have shown remarkable aptitude for improving EE.

In recent years, comprehensive surveys on green cellular networks using various energy saving methods are presented in [4, 12, 31]. Hasan et al. [4] categorized energy saving mechanisms as cooperative networks, adoption of renewable energy resources, deployment of heterogeneous networks, and efficient usage of spectrum. In [31], Xu et al. outlined various distinctive approaches to reduce grid energy consumption in modern cellular networks. The strategies can be broadly classified into energy-efficient hardware design, selectively turning off some components during low-traffic period, optimizing radio transmission process in a physical layer, and powering RANs by renewable energy resources. On the other hand, several research works presented in [13, 32–34] were carried out to improve the EE of the cellular networks with hybrid power supplies. In [13], Peng et al. proposed an energy management technique for cellular networks with the provision of a hybrid energy supply and BS sleep mode. Han and Ansari [33] investigated the optimization of green energy utilization resulting in a significant reduction of conventional grid energy consumption during peak traffic periods. This work did not consider green energy sharing among the BSs. On the other hand, Chia et al. [32] proposed a model for energy sharing between two BSs through a resistive power line, whereas Xu et al. [34] focused on energy harvesting and coordinate transmission-enabled wireless communication by investigating a joint energy and communication cooperative approach. In the proposed paradigm, energy cooperation was implemented by the cellular network operators via signing a contract with the grid operator so that BSs can exchange green energy via the existing grid infrastructure. Besides, the system model has not discussed energy saving issues in this paper. However, none of these papers in [13, 32–34] considered either the JT or the DPS CoMP-based cellular networks for investigating the EE performance or developing energy sharing mechanisms.

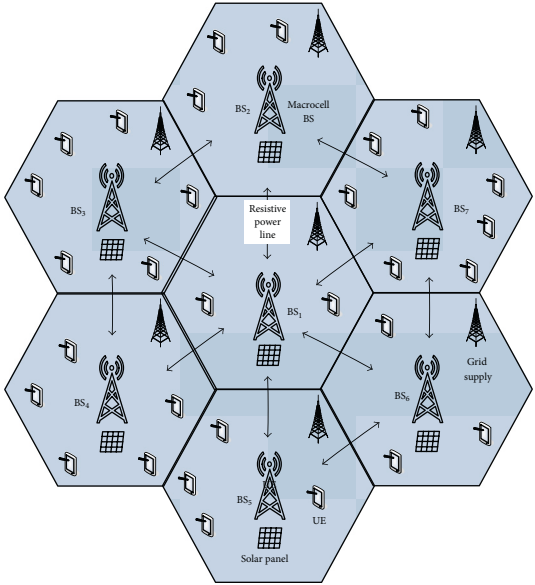

Figure 1: A section of the proposed network model with a hybrid energy supply.

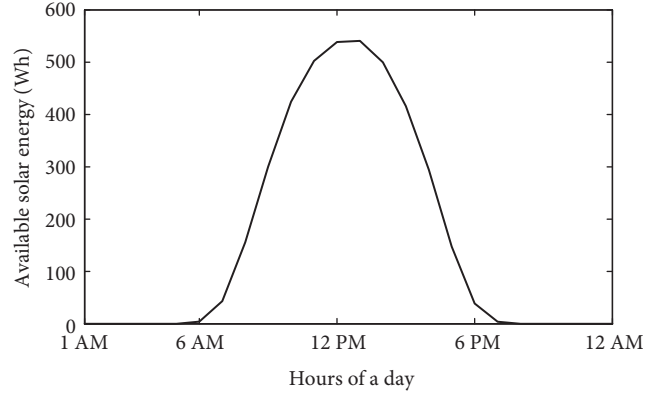

Figure 2: Average hourly solar energy generation.

## 3. System Model

This section presents the proposed network model and other system components in the context of orthogonal frequency division multiple access- (OFDMA-) based LTE-A cellular systems, which can also be adopted to other standards.

*3.1. Network Layout.* The downlink of a multi-cell cellular network having a set of $N$ BSs $\mathbb{B} = \{\mathcal{B}_1, \mathcal{B}_2, \ldots, \mathcal{B}_N\}$ and covering an area $\mathcal{A} = \{\mathcal{A}_1 \cup \mathcal{A}_2 \cup \cdots \cup \mathcal{A}_N\} \subset \mathbb{R}^2$ is considered. Here, $\mathcal{A}_n$ is the coverage area of BS $\mathcal{B}_n$, $n = 1, 2, \ldots, N$. BSs are assumed to be deployed using omnidirectional antennas in a hexagonal grid layout and orthogonal frequency bands are allocated in a BS resulting in zero intracell interference. On the other hand, universal frequency reuse is considered resulting in intercell interference when same frequency band is allocated in two BSs.

All the BSs in the considered LTE-A cellular network are powered by hybrid supplies, namely, PV solar energy and commercial on-grid energy. PV solar energy is the primary energy source, while the grid supply is the standby one. Each BS has an independent on-site solar energy harvester and energy storage device such as a battery bank. For sharing green solar energy among the BSs, each BS is connected with its neighboring BSs through resistive power lines. A segment of the network layout with seven macrocells is depicted in Figure 1. It is also considered that the proposed network is deployed using either the DPS or the JT CoMP transmission technique. On the other hand, UEs are assumed to be distributed uniformly throughout the network. Furthermore, any BS having no user to serve are switched into low power sleep mode for saving energy.

*3.2. Link Model.* This paper considers a channel model with log-normally distributed shadow fading. For a separation $d$ between transmitter and receiver, path loss in dB can be expressed as

$$\text{PL}(d) = \text{PL}(d_0) + 10n\log\left(\frac{d}{d_0}\right), \quad (1)$$

where $\text{PL}(d_0)$ is the path loss in dB at a reference distance $d_0$ and $n$ is the path loss exponent. $\text{PL}(d_0)$ can be calculated using the free-space path loss equation.

Thus, the received power in dBm for $j$th UE at a distance $d = d^{n,j}$ from $n$th BS $\mathcal{B}_n$ is given by

$$P_r^{n,j} = P_t^{n,j} - \text{PL} + X_\sigma, \quad (2)$$

where $P_t^{n,j}$ is the transmitted power in dBm and $X_\sigma$ is the amount of shadow fading modeled as a zero-mean Gaussian random variable with a standard deviation $\sigma$ dB. Then the received SINR $\gamma_{n,j}$ at $j$th UE from $\mathcal{B}_n$ can be given by

$$\gamma_{n,j} = \frac{P_r^{n,j}}{\mathcal{I}_{j,\text{inter}} + \mathcal{I}_{j,\text{intra}} + \mathcal{P}_N}, \quad (3)$$

where $\mathcal{I}_{j,\text{inter}}$ is the intercell interference, $\mathcal{I}_{j,\text{intra}}$ is the intracell interference, and $\mathcal{P}_N$ is the additive white Gaussian noise (AWGN) power given by $\mathcal{P}_N = -174 + 10\log_{10}(\Delta f)$ in dBm with $\Delta f$ as the bandwidth in Hz.

*3.3. Solar Energy Generation Model.* This paper considers the PV solar panel as the on-site green energy harvester. The solar energy generation profile is nondeterministic and depends on some factors, such as temperature, solar light intensity, panel materials, generation technology, and the geographic location of the solar panel. The daily solar energy generation thus shows temporal dynamics over a period of a day in the given area and exhibits spatial variations with geographical location [7]. Due to the tempo-spatial diversity, the available solar energy may not guarantee the adequate energy supplies for a BS to run for a whole day.

Average hourly solar energy generation profile for a full year in Dhaka city of Bangladesh is shown in Figure 2. Here, the solar energy profile for a particular region is estimated by adopting the System Advisor Model (SAM) [35]. The curve indicates that the green energy generation starts from around 6:00 AM, reaches peak value at noon, and stops at about 6:00 PM. SAM supports various solar power generation technologies. However, without losing the generality, distributed type concentrated solar power (CSP) PV technology with 1 kW solar panel is used for generating the shown curve. On the other hand, though solar batteries such as Ni-Cd, NiMH, Li-ion, and sodium nickel chloride are available for using in solar systems, lead-acid batteries are commonly used in solar-powered BSs. The parameters of the solar generation and storage systems for the considered 1 kW solar panel are summarized in Table 1.

3.4. Solar Energy Storage Model. For the proposed system, the green energy storage of the $n$th BS $\mathcal{B}_n$ at time $t$ can be given by

$$s_n(t) = \mu s_n(t-1) + r_n(t) - d_n(t), \quad (4)$$

where $s_n$ is the green energy storage, $r_n$ is the incoming energy from PV solar panel, $d_n$ is the energy demand of the BS, and $0 \leq \mu \leq 1$ is the storage factor, that is, the percentage of storage energy retained after a unit period of time. For example, $\mu = 0.9$ indicates that 10% of energy will be lost in the storage during the time interval. It is to be noted that the stored energy cannot exceed the maximum storage capacity. Therefore, if the generation is higher than the storage capacity, that amount of energy is considered as wastage.

3.5. BS Power Consumption Model. It is important to investigate the traffic demand to be served by the BSs in order to analyze the energy consumption of the network. The mobile traffic volume exhibits both temporal and spatial diversity. Mobile users are assumed randomly distributed. It is also assumed that BSs transmit data to all users with the same data rate. Based on internal surveys on operator traffic data within the EARTH project and the Sandvine report [36], the daily traffic demand in the network is characterized by the normalized traffic profile illustrated in Figure 3.

The BSs energy consumption is directly related to the traffic volumes [37]. The energy consumption of BSs can be subdivided into two parts: the static energy consumption and the dynamic energy consumption. Holtkamp et al. [38] approximated the operating power of a BS as a linear function of RF output power $P_{MAX}$ and BS loading parameter $x$, which can be given by [38]

$$P_{in} = \begin{cases} M_{sec}[P_1 + \Delta_p P_{MAX}(x-1)], & \text{if } 0 < x \leq 1 \\ M_{sec} P_{sleep}, & \text{if } x = 0 \end{cases}, \quad (5)$$

where the expression in the square brackets represents the total power requirement for a transceiver (TRX) chain, $M_{sec}$ is the number of sectors in a BS, and $P_1$ is the maximum power consumption in a sector. The load dependency is accounted for by the power gradient, $\Delta_p$. The loading parameter $x = 1$ indicates a fully loaded system, that is, BS transmitting at full power with all of their LTE resource blocks (RBs) occupied, and $x = 0$ indicates idle state. Furthermore, a BS without any traffic load enters into sleep mode with lowered consumption, $P_{sleep}$. Now $P_1$ can be expressed as follows [38]:

$$P_1 = \frac{P_{BB} + P_{RF} + P_{PA}}{(1 - \sigma_{DC})(1 - \sigma_{MS})(1 - \sigma_{cool})}, \quad (6)$$

where $P_{BB}$ and $P_{RF}$ are the power consumption of baseband unit and radio frequency transceiver, respectively. Losses incurred by DC-DC power supply, main supply, and active cooling can be approximated by the loss factors $\sigma_{DC}$, $\sigma_{MS}$, and $\sigma_{cool}$, respectively. However, power consumption in the power amplifiers is represented by $P_{PA}$ which depends on the maximum transmission power and power amplifier efficiency $\eta_{PA}$ and can be given as follows [38]:

$$P_{PA} = \frac{P_{MAX}}{\eta_{PA}(1 - \sigma_{feed})}. \quad (7)$$

Table 1: Solar panel and storage device parameters.

| Parameters | Type (value) |
|---|---|
| Solar module type | Photovoltaic (distributed) |
| Generation technology | CSP PV cell |
| Solar panel capacity | 1 kWdc |
| DC-to-AC ratio | 0.9 |
| Array type | Fixed roof mount |
| Tilt | 20 degrees |
| Azimuth | 180 degrees |
| Storage type | Lead-acid battery |
| Storage capacity | 2000 Wh |
| Storage factor | 0.96 |

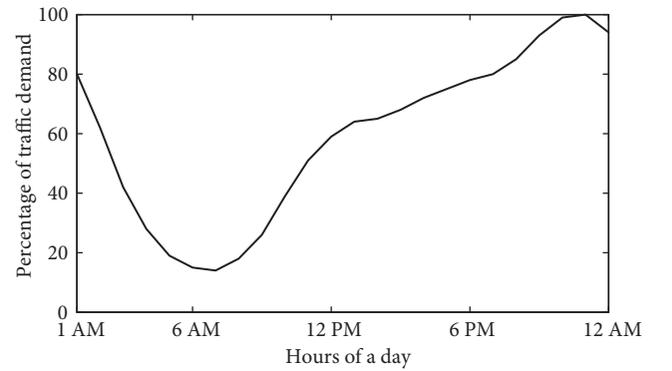

Figure 3: Daily traffic profile of a residential area.

BS power consumption model parameters used in this paper are summarized in Table 2.

3.6. Performance Metrics. This paper evaluates the on-grid energy savings offered by the proposed network models compared to that of the conventional networks powered by grid supply only (i.e., no solar power). The average on-grid energy savings at time $t$ denoted by $E_s(t)$ can be written as

$$E_s(t) = \frac{\sum_{n=1}^{N} P_s(n,t)}{\sum_{n=1}^{N} P_{in}(n,t)} \times 100\%, \quad (8)$$

where $P_{in}(n,t) = d_n(t)$ is the required total power in BS $\mathcal{B}_n$ at time $t$ and $P_s(n,t)$ is the green solar power utilized by the BS $\mathcal{B}_n$ for serving its UEs.

Table 2: BS power consumption model parameters [38].

| Parameters | Value |
| --- | --- |
| BS type | Macro |
| $\eta_{PA}$ | 0.306 |
| $\gamma$ | 0.15 |
| $P_{BB}$ (W) | 29.4 |
| $P_{RF}$ (W) | 12.9 |
| $\sigma_{feed}$ | 0.5 |
| $\sigma_{DC}$ | 0.075 |
| $\sigma_{MS}$ | 0.09 |
| $\sigma_{cool}$ | 0.1 |
| Number of sectors, $M_{sec}$ | 1 |
| Maximum transmit power, $P_{MAX}$ (dBm) | 43 |
| $\Delta_p$ | 4.2 |
| $P_{sleep}$ (W) | 54 |

On the other hand, EE performance metric of a network given in terms of bits per joule can be defined as the ratio of the total throughput to the total power required for running the network. In this paper, we define the EE metric of the proposed network models with CoMP techniques and hybrid power supply as the ratio of the aggregate throughput of the network to that of the net on-grid power consumed by the network. Total achievable throughput in a network at time $t$ can be calculated by Shanon's capacity formula as follows:

$$R_{total}(t) = \sum_{j=1}^{U} \sum_{n=1}^{N_j} \Delta f \log_2(1 + \gamma_{n,j}), \text{ bps}, \quad (9)$$

where $N_j$ is the number of transmitting BSs for serving $j$th UE and $U$ is the total number of UEs in the network. Thus, the EE metric denoted as $\eta_{EE}$ for time $t$ can be written as follows:

$$\eta_{EE}(t) = \frac{R_{total}(t)}{\sum_{n=1}^{N} P_g(n,t)}, \text{ bits/joule}, \quad (10)$$

where $P_g(n,t) = P_{in}(n,t) - P_s(n,t)$ is the on-grid energy consumption in BS $\mathcal{B}_n$ at time $t$.

An alternative performance metric for evaluating the EE of a BS is the Energy Consumption Index (ECI) defined in [39], which can be given by

$$ECI = \frac{P_{in}}{KPI}, \quad (11)$$

where $P_{in}$ refers to the total input power of a BS, whereas KPI (key performance indicator) indicates the total throughput of the BS. In other words, ECI is the reciprocal of EE, and hence, for the proposed networks, it can be evaluated by taking the inverse of (10), while a lower value of ECI implies better EE and vice versa. ECI is more suitable for better visualization of network behavior when the denominator of (10) becomes zero.

## 4. User Association and Algorithm

*4.1. User Association Policy.* The term user association means assigning a UE with a BS for receiving service. Associating users with the closest BS does not always ensure the best SINR due to the randomness of shadow fading. Therefore, user association policy based on the better signal quality (i.e., higher SINR) can support better performance. Therefore, this paper proposes SINR-based user association policy, which is presented for the DPS and JT CoMP-based networks as follows:

(i) *Network with DPS CoMP*. In a network deployed with the DPS CoMP-based transmission technique, one of the available BSs is dynamically selected for serving a UE in the best way. Thus, under the proposed network models with DPS, the BS which provides the highest SINR is selected for associating a UE.

(ii) *Network with JT CoMP*. In JT CoMP-based networks, multiple BSs are dynamically selected for serving a UE. For instance, in a network with 2-BS JT system, two BSs providing the top two SINR values are selected for associating a UE, which jointly transmits data to the particular UE.

*4.2. Energy Sharing Algorithm.* This section presents the proposed energy management scheme with the provision of green energy sharing. Under the proposed network model, each BS is equipped with a PV solar module with a storage facility, which acts as the primary energy source and can be shared among the neighboring BSs. In case there is not adequate energy stored in the storage of a BS, it seeks solar green energy from the neighboring BSs for supporting continuous service for its users. While seeking solar energy, a BS aims to share through the feasible shortest path for minimizing the power loss in the interconnecting resistive transmission line. This implies that a BS can share solar energy only from the six neighboring BSs placed in the first-tier surrounding it as illustrated in Figure 1. Furthermore, a BS aims to share from the BSs having higher solar energy stored.

On the other hand, a neighboring BS shares the surplus energy from its storage only after fulfilling its own demand. If solar energy is not available from the neighboring BSs, only then energy from the standby grid supply is used. Thus, there can arise two different cases for using energy in the BSs, which are presented as below with respect to the $n$th BS $\mathcal{B}_n$.

*4.2.1. Case I: Sufficient Green Energy in Storage.* If $s_n(t) \geq d_n(t)$, then the $n$th BS $\mathcal{B}_n$ has sufficient solar energy for serving its UEs, and hence, the BS will be powered using its own stored energy. Thus, there is no need of green energy sharing from other BSs as well as no on-grid energy is consumed. The remaining solar energy in the storage after fulfilling the demand denoted by $g_n(t)$ can be expressed as

$$g_n(t) = s_n(t) - d_n(t). \quad (12)$$

Therefore, after meeting the demand of time $t$, the available solar energy in the storage of $\mathcal{B}_n$ for the time slot $(t+1)$ can be written as

$$s_n(t+1) = \mu g_n(t) + r_n(t+1) - d_n(t+1), \quad (13)$$

where $r_n(t+1)$ and $d_n(t+1)$ are the generation of solar energy and the total energy demand for time $(t+1)$, respectively.

*4.2.2. Case II: Insufficient Green Energy in Storage.* The scenario with $s_n(t) < d_n(t)$ implies that there is not sufficient solar energy stored for powering the BS $\mathcal{B}_n$, and hence, solar energy sharing is required. Hence, $\mathcal{B}_n$ seeks for the additional solar energy from its neighbors, which is the difference between the total energy demand and the solar energy remaining in its own storage. Therefore, the total green energy required to be shared by $\mathcal{B}_n$ at time $t$ denoted by $g_{n,s}(t)$ can be expressed as

$$g_{n,s}(t) = d_n(t) - s_n(t). \quad (14)$$

For sharing solar energy, BS $\mathcal{B}_n$ sorts its neighbors in a descending order of the available solar energy in their respective storages. Let the set of sorted BSs be given by $\mathbb{B}_n = \{\mathcal{B}_{n,1}, \mathcal{B}_{n,2}, \ldots, \mathcal{B}_{n,M}\}$, where $M$ is the number of neighboring BSs of $\mathcal{B}_n$ for sharing energy and $\mathcal{B}_{n,p}$ has higher storage than $\mathcal{B}_{n,q}$ for $p < q$. Now, if the neighboring BS $\mathcal{B}_{n,1}$ has a shareable solar energy $\geq g_{n,s}(t)$, BS $\mathcal{B}_n$ accepts this amount from $\mathcal{B}_{n,1}$ that fulfills its demand. The sharable solar energy of a BS is the amount that can be shared after fulfilling its own demand. If the sharable energy of $\mathcal{B}_{n,1}$ is $<g_{n,s}(t)$, BS $\mathcal{B}_n$ accepts the amount from $\mathcal{B}_{n,1}$ that it can share.

For the remaining amount of required energy, BS $\mathcal{B}_n$ seeks to share from $\mathcal{B}_{n,2}$ and continues to the next BSs in $\mathbb{B}_n$. Let $\varepsilon_{n,m}$ be the amount of solar energy shared by $\mathcal{B}_n$ from the neighboring BS $\mathcal{B}_{n,m}$. Then the total solar energy received by $\mathcal{B}_n$ from the neighboring BSs can be given by

$$\varepsilon_n(t) = \sum_{m=1}^{M} \alpha_{n,m} \varepsilon_{n,m}(t), \quad (15)$$

where $0 \leq \alpha_{n,m} \leq 1$ is the utilization factor representing the line loss between BS $\mathcal{B}_n$ and its neighbor $\mathcal{B}_{n,m}$, that is, while sharing $\alpha_{n,m} \times 100\%$ of the energy is dissipated as line loss.

If $\varepsilon_n(t) = g_{n,s}(t)$, no on-grid energy is used by BS $\mathcal{B}_n$ at time $t$. Otherwise, on-grid energy is consumed for powering BS $\mathcal{B}_n$ for serving its UEs. Thus, the conventional grid energy consumption denoted by $c_n(t)$ by $\mathcal{B}_n$ at time $t$ can be given by

$$P_g(n,t) = d_n(t) - \varepsilon_n(t). \quad (16)$$

Table 3: Pseudo code of the proposed energy sharing algorithm for $n$th BS $\mathcal{B}_n$.

| | |
|---|---|
| 1: | **Initialize:** $s_n(t), d_n(t), \mu, \alpha_{n,m}, \varepsilon_{n,m} = 0, \forall n = 1, 2, \ldots, N;$ $\forall m = 1, 2, \ldots, M$ |
| 2: | **If** $s_n(t) \geq d_n(t)$ |
| 3: | $P_s(n,t) = d_n(t)$ and $P_g(n,t) = 0$ |
| 4: | $g_n(t) = s_n(t) - d_n(t)$ |
| 5: | **Else** |
| 6: | Coordinate with other BSs for sharing solar energy Sort the neighboring $M$ BSs with respect to stored energy i.e., find the set $\mathbb{B}_n = \{\mathcal{B}_{n,1}, \mathcal{B}_{n,2}, \ldots, \mathcal{B}_{n,M}\},$ $s.t., s_{n,p}(t) \geq s_{n,q}(t)$ for $p < q$ |
| 7: | **For** $m = 1 : M$ |
| 8: | Calculate $r_{n,s}(t) = g_{n,s}(t) - \sum_{k=0}^{m-1} \alpha_{n,k} \varepsilon_{n,k}(t),$ $\varepsilon_{n,k}(t) = 0$ for $k < 1$ |
| 9: | **If** $s_{n,m}(t) - d_{n,m}(t) \geq r_{n,s}(t)$ |
| 10: | Share solar energy $\varepsilon_{n,m}(t) = r_{n,s}(t)$ from $\mathcal{B}_{n,m}$ |
| 11: | **Else** |
| 12: | Share solar energy $\varepsilon_{n,m}(t) = s_{n,m}(t) - d_{n,m}(t)$ from $\mathcal{B}_{n,m}$ |
| 13: | **If** $\sum_{k=1}^{m} \alpha_{n,k} \varepsilon_{n,k}(t) = g_{n,s}(t)$ |
| 14: | Stop the algorithm and Go to Step 21 |
| 15: | **Else** |
| 16: | $m = m + 1$ and Go to Step 8 |
| 17: | **End If** |
| 18: | **End If** |
| 19: | **End For** |
| 20: | **If** $\sum_{m=1}^{M} \alpha_{n,m} \varepsilon_{n,m}(t) = g_{n,s}(t)$ |
| 21: | $P_s(n,t) = d_n(t)$ and $P_g(n,t) = 0$ |
| 22: | Stop the algorithm |
| 23: | **Else** |
| 24: | $P_s(n,t) = \varepsilon_n(t) = \sum_{m=1}^{M} \alpha_{n,m} \varepsilon_{n,m}(t)$ |
| 25: | $P_g(n,t) = d_n(t) - \varepsilon_n(t)$ |
| 26: | **End If** |
| 27: | **End If** |

Pseudo codes of the energy sharing algorithm with respect to the $n$th BS $\mathcal{B}_n$ is presented in Table 3. In the pseudo code, $d_{n,m}(t)$ and $s_{n,m}(t)$ are the total energy demand and the stored solar energy of the $m$th BS $\mathcal{B}_{n,m} \in \mathbb{B}_n$.

## 5. Performance Analysis

*5.1. Simulation Setup.* This section analyzes the performance of the proposed cellular network framework with BSs powered by PV solar energy and standby grid supply. A MATLAB-based Monte Carlo simulation platform is developed for carrying out extensive simulations. For each data point, results are calculated by averaging over 10,000 independent iterations each with a simulation time of seven days. The network is deployed using a hexagonal grid layout with a cell radius of 1000 m. For comprehensive performance evaluation, intercell interference contributions by the 18 BSs placed in the two surround tiers are taken into consideration. On the other hand, UEs are considered uniformly distributed over the geographical area. Performance of the proposed network model for any nonuniform UE distribution can also be evaluated in a similar way. It is assumed that one UE occupies one RB and equal transmit power over all RBs. Furthermore, under the JT CoMP-based network, it is considered that the top two BSs providing the best SINR values serve a UE simultaneously. A summary of the system parameters of the simulated network are set in reference to the LTE standard [18] as summarized in Table 4. On the other hand, unless otherwise specified, the proposed network models are simulated considering $x$ as a uniform random variable in [0,1] for modeling the spatial variation in traffic generation among the BSs, while equal solar generation is assumed in all the BSs having the solar module parameters as presented in Table 1.

### 5.2. Result Analysis

*5.2.1. SINR and Throughput Analysis.* Figure 4 demonstrates the empirical cumulative distribution function (CDF) of received SINR at UEs located throughout the considered network model. The network is considered with fully loaded (i.e., $x = 1$) BSs, which are supplied with hybrid supply with no inter-BS energy sharing. From the figure, a clear distinction in SINR distribution is observed among the JT CoMP-, DPS CoMP-, and non-CoMP-based hybrid systems. The JT CoMP-enabled hybrid system keeps its optimistic nature achieving comparatively stronger SINR among UEs, which ranges around $-2$ dB to 60 dB. This is because the two BSs offering the highest SINR values simultaneously serve a UE resulting in significantly better signal quality. In contrast, a non-CoMP-based hybrid scheme has the worst SINR performance as it spreads out over a larger range compared to that of the other techniques. SINR performance of the DPS CoMP-based system lies in between the JT and the non-CoMP-based systems as it selects the BS supporting the best signal quality. Thus, from the view point of SINR, the JT CoMP-based hybrid model is a preferred choice compared to the others.

On the other hand, a comparison of the throughput performance over a day among the different hybrid schemes is shown in Figure 5. As seen from the figure, throughput curves clearly follows the given traffic pattern (Figure 3). Also, a clear distinction is observed between the low traffic and the peak traffic times. This is because higher traffic means the allocation of higher number of RBs to the users resulting in higher throughput. Further, during peak traffic arrivals, throughput gap is more significant among the different hybrid systems. On the other hand, it is observed that the JT CoMP-based hybrid system outperforms others in terms of throughput. As throughput is directly related to the signal quality, superior SINR performance of the JT CoMP-based

TABLE 4: Simulation parameters.

| Parameters | Value |
|---|---|
| Resource block (RB) bandwidth | 180 kHz |
| System bandwidth, BW | 10 MHz (50 RBs), 600 subcarriers |
| Carrier frequency, $f_c$ | 2 GHz |
| Duplex mode | FDD |
| Cell radius | 1000 m |
| BS transmission power | 43 dBm |
| Noise power density | $-174$ dBm/Hz |
| Number of antennas | 1 |
| Reference distance, $d_0$ | 100 m |
| Path loss exponent, $n$ | 3.574 |
| Shadow fading, $\sigma$ | 8 dB |
| Access technique, DL | OFDMA |
| Traffic model | Randomly distributed |

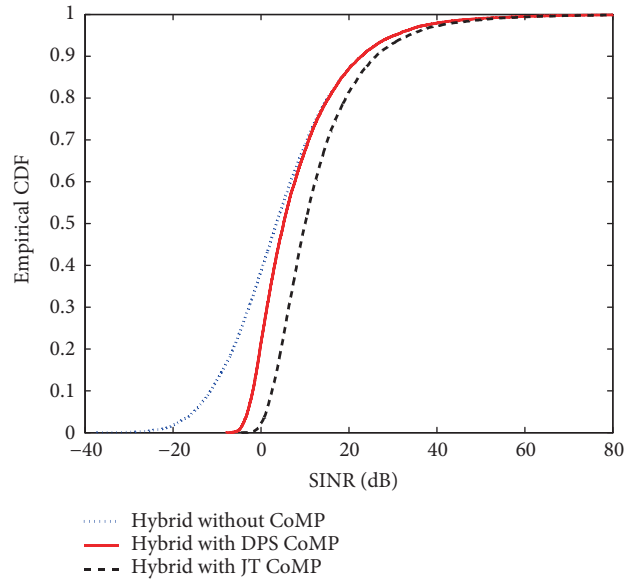

Figure 4: Empirical CDF of received SINR among the different hybrid systems for $x = 1$.

model as observed in Figure 4 ensures its relatively higher throughput.

*5.2.2. Energy Analysis with Hybrid Supply under No Energy Sharing.* Figure 6 compares the temporal variation of energy consumption by a BS in a conventional cellular system to that in the proposed network model with hybrid energy supply without energy sharing option. Here, the conventional scheme implies a cellular system powered by only grid energy with no energy sharing option and no CoMP transmission mechanism. As seen from the figure, on-grid consumption of a BS in the conventional system follows traffic distribution, reaches at peak when traffic demand is the highest, and goes down as the traffic decreases. Use of solar energy in BSs and sharing this energy among BSs is to minimize this grid energy usage. The energy consumption curve for the hybrid model with no sharing option is also presented assuming a storage capacity of 2 kWh. As seen, up to around 5 AM, a BS is completely run by on-grid supply as solar energy is unavailable during this period. After this, on-grid energy consumption gradually decreases with the increase of solar energy availability and becomes zero at 9 AM. Between 9 AM and 9 PM, there is adequate solar energy available for running the BS, and hence, no consumption of conventional energy. During this period, a BS fulfill its demand from its own solar energy storage and stores the surplus energy for future use. As time goes, stored solar energy decreases gradually with the decrease of solar light intensity, and once again, on-grid energy is required to serve its associated users after 9 PM. Thus, due to this temporal dynamics of solar energy generation, the available solar energy is not always sufficient for supplying the BS, and hence, on-grid energy is still required to fully meet the BS demand.

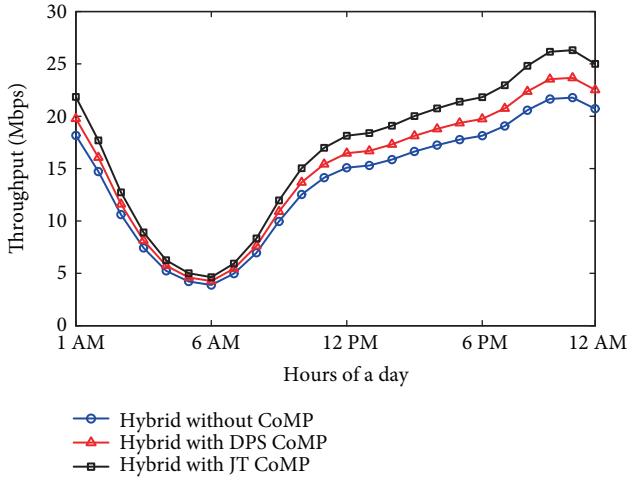

Figure 5: Throughput comparison of a single BS among the different hybrid models.

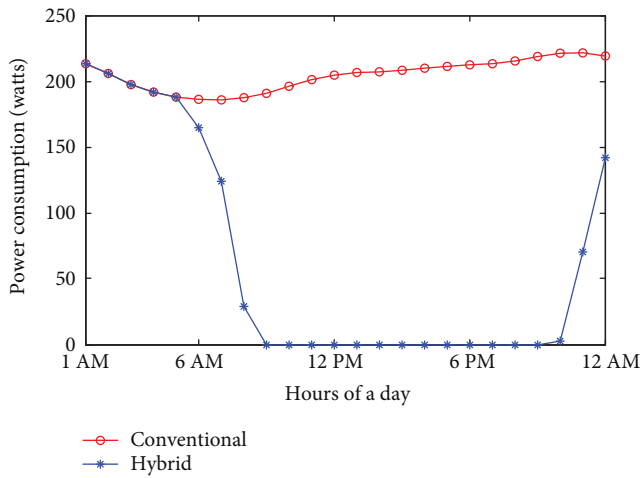

Figure 6: Comparison of on-grid power consumption in a single BS between the existing hybrid energy model (no sharing) and the conventional scheme with grid supply only. No CoMP technique is implemented in the network models.

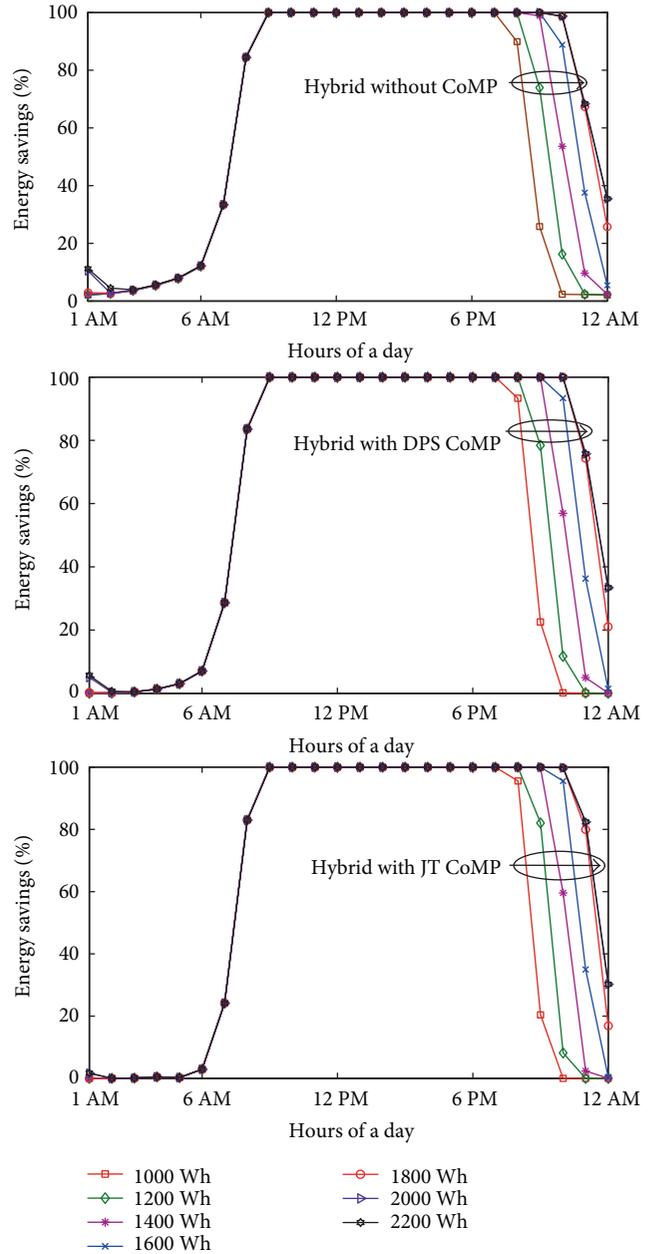

Figure 7: Average on-grid savings of a BS for different storage capacities with 1 kW solar module and no energy sharing.

Percentage of energy savings under the different network models with hybrid power supply for various solar storage capacity is demonstrated in Figure 7. Here the network model having hybrid energy supply with no energy sharing and non-CoMP transmission refers to the existing hybrid system. As seen from the figure, during 12 AM to 6 AM, the BSs are mainly run by grid energy, and hence the savings in this period are mainly due to the switching of some BSs having no traffic into sleep mode. Under JT CoMP, energy savings in this period are found almost zero as the probability of BSs to enter into sleep mode is negligible. As the time proceed from 6 AM, solar energy generation increases leading to higher on-grid energy savings, which eventually reaches the peak around noon and following a gradual decrease around the evening. Thus, the hybrid scheme has the potential to reduce the on-grid energy consumption up to 100% for a prolonged period of time as evident from the figure. On the other hand, a significant impact of storage capacity on the energy savings is observed. As seen from the figure, the energy savings region is expanded with the increase of storage capacity resulting in higher savings. For example, energy savings curve for 2 kWh storage capacity lasts longer compared to that for a capacity of 1 kWh. Furthermore, the saving curves for 2 kWh is fully overlapped with that of 2.2 kWh and further increases in the storage capacity have no impact on energy savings as evident from the figure. Thus, the optimal value of storage capacity is 2 kWh. Notably, the energy saving performance follows similar fashion for all the three hybrid models having no significant variation due to CoMP techniques. This implies that for the particular network setting, savings is dependent predominantly on the solar energy generation and storage capacity.

A comparison of ECI performance metric with the three hybrid models is presented in Figure 8. During the low traffic periods in the morning, ECI increases rapidly up to a certain point and then starts to fall beyond that point. The upward trending nature of ECI implies the relatively higher on-grid energy consumption as the available solar energy is almost negligible. With the increase of solar energy generation, the ECI curve is pushed downward. It can be seen that during 9 AM to 9 PM, the ECI curve falls to zero indicating no on-grid energy consumption, that is, maximum EE. With the diminishing sunlight, stored energy also runs out by powering the BSs, and once again after 9 PM, on-grid energy is required to supply the BSs resulting in an upward trend of the ECI curve. On the other hand, since throughput performance of JT CoMP is better than that of DPS CoMP as observed in Figure 5, the JT CoMP-based hybrid system provides superior ECI performance. It can also be seen that the proposed network models have better ECI performance compared to that of the existing hybrid system.

Figure 9 illustrates the variation of average EE and the average on-grid power consumption in a BS under different hybrid schemes with the solar storage capacity. The network is simulated over a period of a week considering 1 kW solar panel in each BS as a green energy harvester. The three power consumption curves decreases in a similar fashion. The tendency of the down trending of the energy consumption curves indicates that the energy drawn from the grid decreases with the increment of storage capacity. The curves eventually reach to their respective constant values, which is also supported by Figure 7. This is because the storage limit reached optimum value for the given solar panel capacity and further increase in storage capacity does not make any significant improvement in reducing on-grid consumption. It is also observed that the grid power consumption is slightly higher under the JT CoMP-based hybrid system as two BSs simultaneously serve a particular UE. On the other hand, the EE curves demonstrate the opposite trend with the increment of storage capacity. However, the JT-based hybrid model shows superior EE performance compared to the DPS-based and the existing non-CoMP-based hybrid model due to its significantly higher throughput. As expected, EE metric also eventually reaches its peak and remains constant after the optimal storage capacity, which is found equal to 2 kWh.

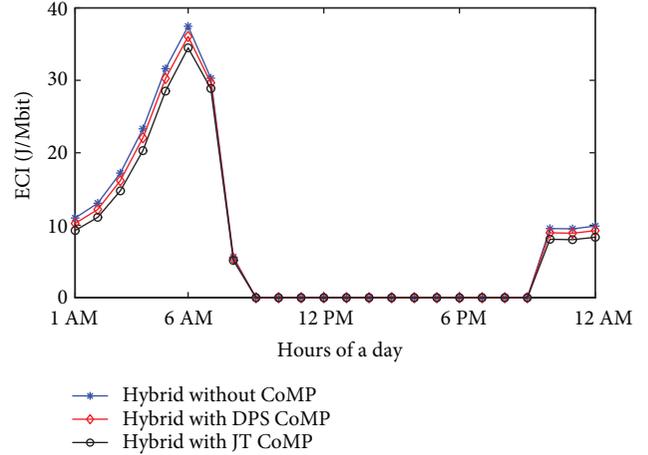

Figure 8: ECI comparison among the different hybrid models without sharing.

#### 5.2.3. Energy Analysis with Hybrid Supply under Energy Sharing

*(1) Equal Solar Generation Capacity.* Figure 10 presents the impact of the resistive loss in the green energy transmission lines on the EE metric under different hybrid models evaluated over a period of one week. The systems are simulated assuming that each BS has an equal solar panel capacity of 1 kW with the optimal storage capacity of 2 kWh. Results for the three different scenarios, namely, hybrid only with no CoMP, hybrid with DPS CoMP, and hybrid with JT CoMP, are shown. From the figure, a clear difference is noticed in the EE metric performance with energy sharing and no sharing techniques of the respective hybrid models. The figure depicts that the EE performance has a decreasing trend with the increase of line loss. For the case of the hybrid model with no CoMP system, up to a certain percentage of line loss (around 55%), the EE performance with energy cooperation among BSs remains better than that of the corresponding non-cooperation-based scheme. Beyond this amount of line loss, energy cooperation degrades the network EE. Further analysis of the figure identifies that the concept of solar energy sharing can improve the EE of the DPS-based CoMP scheme if the line loss is less than 15%, while apparently no positive impact of such cooperation is observed for the JT-based CoMP scheme.

Dependency of EE on the storage factor of batteries over a period of a week is illustrated in Figure 11. The solar panel capacity and the storage capacity are same as those of Figure 10. As shown, all of the curves have a similar pattern reaching their respective peak values at a storage factor of 1. Storage factor indicates percentage of storage energy retained after unit period of time and the higher value of $\mu$ provides better EE. For the hybrid case with no CoMP mechanism, energy sharing demonstrates a positive impact on EE with the increase of storage factor $\mu$. However, the best EE is found for the JT-based hybrid model in which the curves of sharing and no sharing are fully overlapping with each other, whereas the EE curves of DPS CoMP lie in between the former two hybrid models.

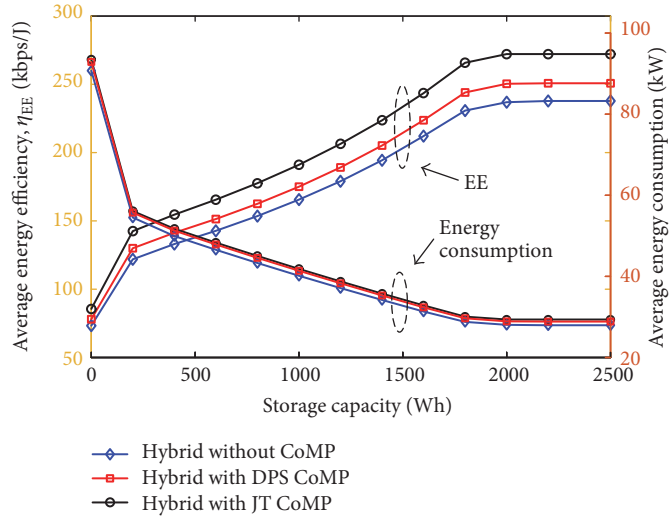

Figure 9: Comparison of EE and on-grid power consumption for the three different hybrid schemes without sharing.

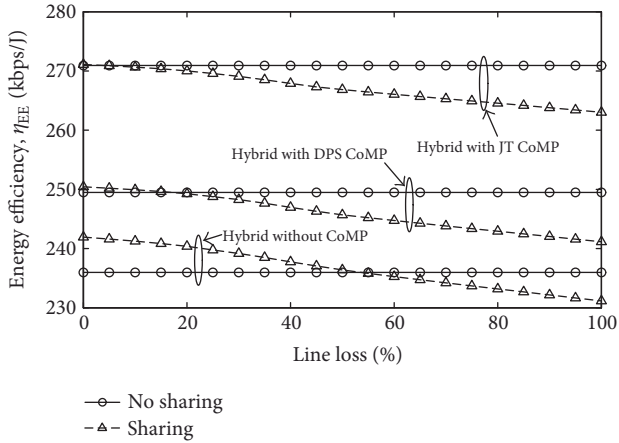

Figure 10: EE variation with line loss under the different hybrid scenarios with energy sharing considering equal solar capacity.

On the other hand, Figure 12 illustrates the impact of solar panel capacity on the EE performance of the proposed models. With the increase of solar panel capacity, storage capacity is also linearly scaled for guaranteeing no wastage of generated solar energy. As evident from the figure, with the increase of solar capacity (i.e., higher available solar energy), EE of the proposed network models substantially improves, which is mainly due to the increasing use of solar energy for running the BSs. From the figure, we can also determine the minimum solar capacity required for running the BSs 24 hours from solar energy. This is the capacity beyond which EE becomes infinity implying that no on-grid energy is required. As seen in the figure, this capacity is found around 1800 W and 1700 W for no sharing and sharing schemes, respectively. Once again, the difference between the energy sharing models and the corresponding no sharing models is not that much significant as also observed in Figures 10 and 11. Furthermore, comparison of Figures 10–12 clearly demonstrates that the proposed hybrid models with CoMP techniques and energy sharing mechanism has superior EE performance compared to that of the existing hybrid system.

*(2) Spatial Diversity in Solar Energy Generation.* On top of the temporal diversity in solar generation in a BS as considered in the previous simulations, this section presents the results by introducing the spatial diversity as well. This phenomenon of spatial diversity in solar energy generation among the BSs is modeled as a uniform random variable distributed in [0,1] multiplied by a constant $c_s$. Unless otherwise specified, $c_s = 1$ kW is used for the simulations.

Under such scenario, comparison of EE with the resistive line loss evaluated over one week is illustrated in Figure 13. The figure follows the similar fashion of Figure 10. However, the figure depicts significant performance gap between the energy sharing and the corresponding no energy sharing based schemes. As the solar generation now varies from BS to BS, energy cooperation becomes more effective for improving EE by sharing surplus solar energy of some BSs with other BSs having lower amount. For the same reasons, compared to Figure 11, Figure 14 demonstrates a clear impact of energy sharing on the EE with varying storage factor $\mu$. Furthermore, EE gap between the energy sharing and the corresponding no sharing cases is found higher for lower values of $\mu$, which diminishes as $\mu$ increases to 1. The lower storage factor indicates lower amount of useful energy stored in the batteries, which stimulates the necessity of sharing solar energy from the neighboring BSs leading to higher gap and vice versa.

Figure 15 presents the variation of EE with the solar module capacity demonstrating the impact of energy sharing and no sharing operation for a network with spatial diversity of solar energy generation. Any capacity shown in the $x$-axis, for instance, 1 kW, implies that the solar energy generation varies among the BSs according to a uniform random variable distributed in [0,1] with $c_s = 1$ kW. As expected, EE of the proposed networks improves with the increase of solar capacity. Furthermore, comparison with the Figure 12 identifies that the positive impact of energy sharing on the EE is more apparent under the case of varying solar energy generation, which can be explained in the same way as presented for Figures 13 and 14. Once again, EE performance of the proposed CoMP-based systems are found significantly better than that of the existing hybrid system as illustrated in Figures 13–15.

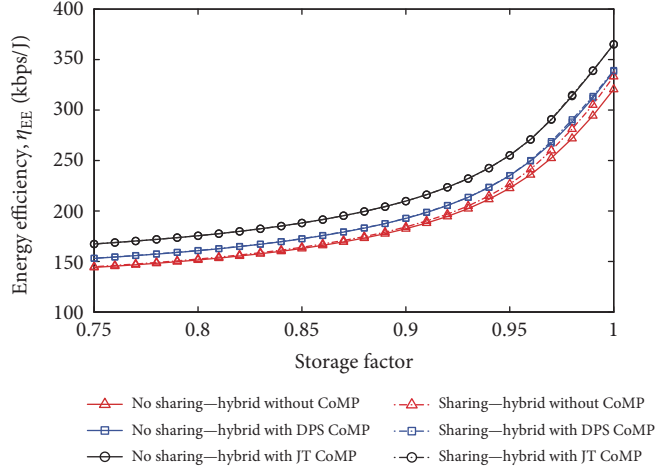

Figure 11: Variation of EE with storage factor with energy sharing considering equal solar capacity.

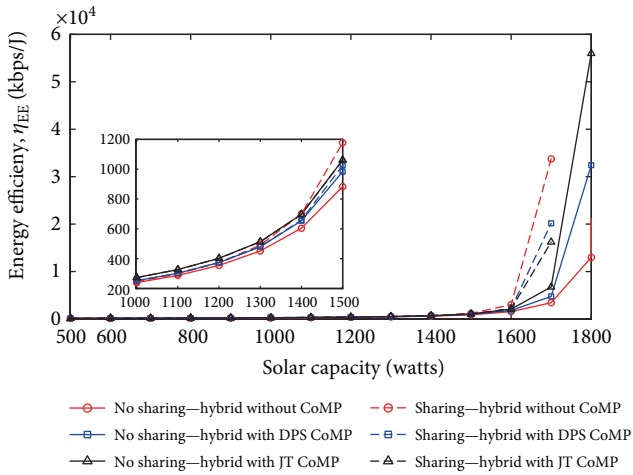

Figure 12: EE of the different hybrid schemes with energy sharing under equal solar capacity.

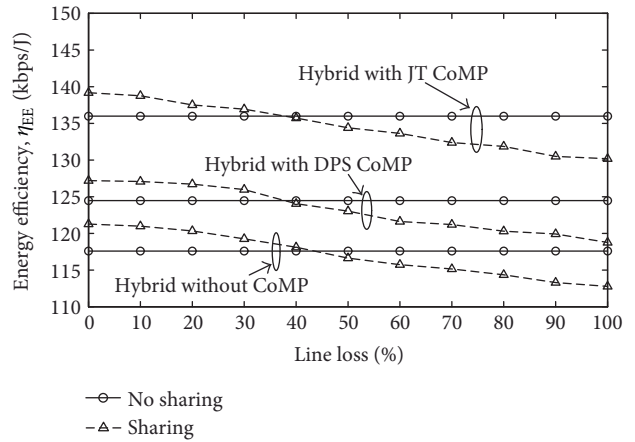

Figure 13: EE with line loss for different hybrid scenarios with energy sharing under varying solar capacity.

## 6. Conclusion

This paper have proposed a framework for an energy efficient cellular network with hybrid-powered BSs, where a PV solar module acts as the main energy source for a BS and the grid remains as the standby. A solar energy sharing algorithm among the BSs has also been proposed for further greening the cellular networks by minimizing on-grid energy consumption. The proposed framework has been analyzed for both the DPS and the JT CoMP transmission techniques based on future cellular networks. System performance has been evaluated in terms of EE, energy savings, and throughput by comprehensive Monte Carlo simulations under varying system parameters, such as storage capacity, resistive line loss, storage factor, solar generation capacity, and CoMP techniques. Simulation results have shown that EE and energy savings of the proposed hybrid system increase with the increase in the storage capacity but approach to a peak value after a certain optimum capacity beyond which no further improvement is inflicted. Moreover, a continuous increase in EE has been observed for better storage factors, whereas the resistive loss in the transmission lines has been found to have significant deteriorating impact resulting in reduced improvement in EE. On the other hand, the proposed solar energy cooperation among the neighboring BSs have demonstrated a significant improvement in network EE for all the CoMP- and non-CoMP-based networks. It has also been identified that the JT CoMP hybrid system has the best EE performance compared to the DPS CoMP scheme. Moreover, the proposed network models have always been found higher energy efficient than the existing hybrid scheme with no energy sharing and non-CoMP transmission. Furthermore, energy sharing has been figured out more effective for improving EE in networks having spatial diversity in solar energy generation. In summary, the degree of improvement in EE of the proposed hybrid-powered network models with and without the proposed energy sharing mechanism has been found highly dependent on the network scenarios.

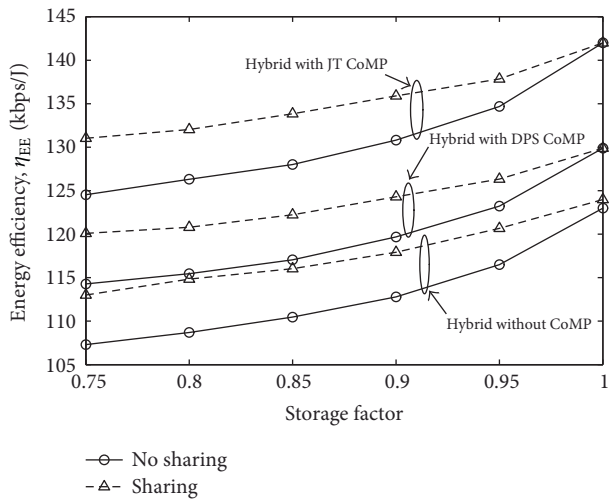

Figure 14: EE with storage factor for different hybrid scenarios with energy sharing under varying solar capacity.

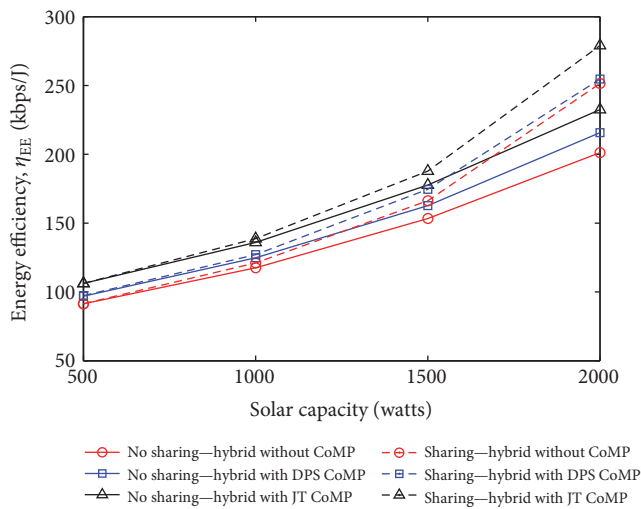

Figure 15: EE for different hybrid schemes with energy sharing under varying solar capacity.

## Conflicts of Interest

The authors declare that there is no conflict of interest regarding the publication of this paper.